\documentclass[aps,prl,twocolumn,groupedaddress]{revtex4}
\usepackage{longtable}
 
\begin{document}
 
\preprint{LA-UR-08-0043}
 
\title{Multiple Extremal Eigenpairs of Very Large Matrices by Monte Carlo Simulation}
 
 
\author{T. E. Booth}
\affiliation{ Applied Physics Division, Los Alamos National Laboratory, Los Alamos, NM 87545}
\author{J. E. Gubernatis}
\affiliation{Theoretical Division, Los Alamos National Laboratory, Los Alamos, NM 87545}

 
\date{\today}

\begin{abstract}
We present a new Monte Carlo algorithm that allows the simultaneous determination of a few extremal eigenpairs of a very large matrix. It extends the power method and uses a new sampling method, the sewing method, that does a large state space sampling as a succession of samplings from a smaller state space. We illustrate the new algorithm by its determination of the two largest eigenvalues of the transfer matrix of a square Ising model at the critical temperature for sizes from $16\times 16$ to $48\times 48$. 
\end{abstract}

 
\maketitle
 
 
A common problem in computational physics is computing the eigenpairs of large matrices. We will present a new Monte Carlo algorithm that allows the simultaneous determination of a few extremal eigenpairs of a very large matrix without the need to orthogonalize pairs of vectors to each other or store all the components of any vector. 

The new algorithm is an extension of the power (projection) method \cite{wilkinson}, which is the traditional starting point for a Monte Carlo determination of the eigenpair associated with the eigenvalues of largest absolute value $\lambda_1$. While various Monte Carlo versions of the power method often compute this dominant eigenvalue very well, computing subdominant eigenvalues $ \lambda _2 ,\lambda _{3,} \ldots $ has often proven much more difficult and is much less frequently attempted \cite{hammond}. Our Monte Carlo power method computes multiple extremal eigenpairs simultaneously and straightforwardly. For the particular algorithm presented we also introduce a new sampling method, the sewing method, that does a large state space sampling as a succession of small state space samplings. Although our new algorithm is extendable to finding more than two extremal eigenpairs, here we will focus on finding just $ \lambda _1$ and $\lambda _{2}$.

Our ultimate targets are matrices so large that they are unassailable deterministically because no single vector can be stored in memory. As the system size increases, finding a few extremal eigenpairs of the transfer matrix of the two-dimensional Ising model becomes such a problem. Its two extremal eigenvalues are also of significant physical interest: the logarithm of $\lambda _1 $ is proportional to the free energy, and the ratio $ \lambda _2 /\lambda _{1} $ controls long range spin correlations near the critical point \cite{thompson,onsager}.  Onsager \cite{onsager} derived exact expressions the eigenvalues of this transfer matrix for any finite-sized system thereby providing a nearly unprecedented opportunity for non-trivially benchmarking our algorithm for exceptionally large systems. We comment however that there is no {\it a priori\/} restriction of our algorithm to problems in classical statistical mechanics. It is also applicable to transfer matrices of quantum origin and to ground and low lying excited states of many Hamiltonian matrices. In addition it has application to more widely diverse problems as the nuclear critically problem \cite{lux}. We also note that our extension of the power method is not limited to a Monte Carlo implementation \cite{gubernatis}.
 
The transfer matrix $A(\sigma,\sigma')$ of an $m\times m$ Ising model with periodic boundary conditions in one direction and open boundary conditions in the other is a $2^m \times 2^m$ matrix whose elements are \cite{thompson}
\begin{equation}
A\left( \sigma ,\sigma'\right) = \exp \left(\nu \sum\limits_{k = 1}^{m -1}
\mu _k \mu _{k + 1} \right)  \exp \left(\nu \sum\limits_{k = 1}^m
\mu _k \mu _k^{'} \right)
\label{eq:transfer_matrix}
\end{equation}
with  $\nu=J/k_B T$, $J$ is the exchange constant, $k_B$ being Boltzmann's constant, and $T$ equal to the temperature. The Ising spin variable $\mu_{k}$ has the value of $ \pm 1$, $\mu _{m + 1}  = \mu _{1}$, and the symbol $\sigma = \left( {\mu _{1} ,\mu _{2} , \ldots ,\mu _{m} } \right)$ denotes a configuration of Ising spins.  (There are $2^m$ possible configurations .)  Numerically, we represented a $\sigma$ by the first $m$ bits of integers ranging from 0 to $2^m-1$.  We comment that all the elements of $A(\sigma,\sigma')$ are greater than zero so the matrix is maximally dense, and because of the one open boundary, it is also asymmetric. From the Perron-Frobenius Theorem \cite{wilf} we have that a dominant eigenvalue that is real and positive. Further, all components of the corresponding eigenstate are real and have the same sign.

The power method \cite{wilkinson} for some real-valued $M\times M$ matrix $A$, not necessarily symmetric, is an iterative procedure, started with some  $\psi$, normalized in a convenient but otherwise relatively arbitrary, manner, that cycles the two steps
\begin{equation}
\begin{array}{c}
 \phi = A\psi  \\
 \psi = \phi / \| \phi \| \\
 \end{array}
\label{eq:power_method}
\end{equation}
until some convergence criterion is met. If $(\lambda_i,\psi_i)$ are eigenpairs of $A$, then starting with some
\[
\psi= \sum\limits_{\alpha = 1}^M {\omega_\alpha \psi_\alpha }
\]
and specifying $\left| {\lambda _1 } \right| > \left| {\lambda _2 } \right| \ge \left|{\lambda _3 } \right| \ge \cdots  \ge \left| {\lambda _N } \right|$, we find after $n$ iterations that
\begin{equation}
A^n\psi  = \lambda _1^n \left[ {\omega_1 \psi_1  + \sum\limits_{\alpha= 2}^M {\omega_\alpha \left( {\frac{{\lambda _\alpha}}{{\lambda _1 }}} \right)^n \psi_\alpha} } \right]
\label{eq:pm}
\end{equation}
Accordingly, $\psi  \to \psi_1 /\|\psi_1\| $ and $\|\phi\| \to \lambda _1 $ as $n\rightarrow \infty$.

Finding more than one eigenpair by the power method requires initializing the method with more than one starting point \cite{wilkinson,golub}. For methods of which we are aware, these starting points need to be orthogonal, and this orthogonality needs to be maintatined, at least periodically, throughout the iteration. This is much more difficult to do in a Monte Carlo procedure than in a determinisitic one \cite{hammond}. 

In developing a Monte Carlo algorithm to estimate two extremal eigenvalues, we will exploit several observations of Booth \cite{booth1,booth2}. He noted that for any eigenpair $(\lambda,\psi)$ and for each non-zero {\it component\/} of the eigenvector, the eigenvalue equation $A\psi  = \lambda\psi$ can be rewritten as
\begin{equation}
\lambda  = \frac{{\sum\limits_j  {A_{i j } \psi _j  } }}{{\psi _i  }}
\label{eq:eigenvalue_estimator}
\end{equation}
and that similar equations can also be written for any number of groupings of components,
\begin{equation}
\lambda  = \frac{{\sum\limits_{i  \in R_1 } {\sum\limits_j  {A_{i j } \psi _j  } } }}{{\sum\limits_{i  \in R_1 } {\psi _i  } }} 
= \frac{{\sum\limits_{i  \in R_2 } {\sum\limits_j  {A_{i j } \psi _j  } } }}{{\sum\limits_{i  \in R_2 } {\psi _i  } }} 
=  \cdots  
= \frac{{\sum\limits_{i  \in R_L } {\sum\limits_j  {A_{i j } \psi _j  } } }}{{\sum\limits_{i  \in R_L } {\psi _i  } }}
\label{eq:groupings}
\end{equation}
where the $R_i$ are rules for different groupings.  Here, we will exploit the fact  that any two groupings, say 1 and 2, imply
\begin{equation}
\sum\limits_{i  \in R_2 } {\psi _i  } \sum\limits_{i  \in R_1 } {\sum\limits_j  {A_{i j } \psi _j  } }  = \sum\limits_{i  \in R_1 } {\psi _i  } \sum\limits_{i  \in R2} {\sum\limits_j  {A_{i j } \psi _j  } } 
\label{eq:cross_product}
\end{equation}
This directly follows from (\ref{eq:groupings}).

As do standard procedures for finding for just the two extremal eigenvalues, we also use two normalized, starting points $\psi'  = \sum\nolimits_\alpha {\omega^\prime_\alpha \psi _\alpha } $ and $\psi'' = \sum\nolimits_\alpha {\omega^\prime _\alpha \psi _\alpha } $ \ but they need not necessarily be orthogonal \cite{booth1,booth2,gubernatis}.  At each step of the power method, we apply $A$ to them individually; however,  to prevent both from projecting to the same dominant eigenfunction, we adjust at each step the relationship between their {\it sum} to direct one to the dominant state and the other to the next dominant one. We do this in the following way: we start the iteration with $\psi=\psi'+\eta\psi''$. If at the $n^{th}$ step, $\psi'$ and $\psi''$ have iterated to $\hat\psi'$ and $\hat\psi''$, then at the $(n+1)^{th}$ step, we use (\ref{eq:groupings}) and (\ref{eq:cross_product})
\begin{eqnarray}
\lefteqn{\frac{{\sum\limits_{i  \in R_1 } {\sum\limits_j  {A_{i j } \hat \psi '_j  } }  + \eta \sum\limits_{i  \in R_1 } {\sum\limits_j  {A_{i j } \hat \psi ''_j  } } }}{{\sum\limits_{i  \in R_1 } {\hat \psi '_i  }  + \eta \sum\limits_{i  \in R_1 } {\hat \psi ''_i  } }}  }\\ \nonumber
&=& \frac{{\sum\limits_{i  \in R_2 } {\sum\limits_j  {A_{i j } \hat \psi '_j  } }  + \eta \sum\limits_{i  \in R_2} {\sum\limits_j  {A_{i j } \hat \psi ''_j  } } }}{{\sum\limits_{i  \in R_2 } {\hat \psi '_i  }  + \eta \sum\limits_{i  \in R_2 } {\hat \psi ''_i  } }}
\label{eq:balance}
\end{eqnarray}
to obtain
\begin{equation}
q_2\eta^2+q_1\eta+q_0=0
\label{eq:quadratic}
\end{equation}
The algorithm thus is to apply $A$ repeatedly to $\psi'$ and $\psi''$. If two real solutions $\eta_1$ and $\eta_2$ of (\ref{eq:quadratic}) exist, then update via
\begin{eqnarray}
\psi' &\leftarrow& A\psi'+\eta_1A\psi'' \\ \nonumber
\psi'' &\leftarrow&A \psi'+\eta_2A\psi''
\end{eqnarray}
otherwise via
\begin{eqnarray}
\psi' &\leftarrow& A\psi'\\ \nonumber
\psi'' &\leftarrow& A\psi''
\end{eqnarray}
After the $\eta$'s become real, $\eta_1$ guides further iterations to $(\lambda_1,\psi_1)$; $\eta_2$, to $(\lambda_2,\psi_2)$. The eigenvalues are estimated from
\begin{eqnarray}
\lambda_1  &=& \frac{ \sum\limits_{i\in R_1}\sum\limits_j A_{ij}\psi_j'
              +\eta_1 \sum\limits_{i\in R_1}\sum\limits_j A_{ij}\psi_j'' }
                    { \sum\limits_{i\in R_1}\psi_i'
               +\eta_1\sum\limits_{i\in R_1}\psi_i'' }\nonumber \\
\lambda_2  &=& \frac{ \sum\limits_{i\in R_1}\sum\limits_j A_{ij}\psi_j'
              +\eta_2 \sum\limits_{i\in R_1}\sum\limits_j A_{ij}\psi_j'' }
                    { \sum\limits_{i\in R_1}\psi_i'
               +\eta_2\sum\limits_{i\in R_1}\psi_i'' }
\label{eq:eigenvalues}
\end{eqnarray}
where $\eta_1$ and $\eta_2$ generate the largest and next largest eigenvalue estimates. A justification of this procedure is given in \cite{gubernatis}.

The Monte Carlo method is used to estimate the result of the repeated matrix-vector multiplication.  In the basis defining the matrix elements of $A$, we write
\[
\begin{array}{c}
 \psi'  = \sum\limits_i {\omega_i ' } \left| i \right\rangle  \\ 
 \psi'' = \sum\limits_i {\omega_i ''} \left| i \right\rangle  \\ 
 \end{array}
\]
and call the amplitudes $\omega'_i$ and $\omega''_i$ {\it weights\/} even though they are not necessarily all positive nor are the sums of their absolute values unity. We assume that the elements of the $M\times M$ matrix $A$ are easily generated on-the-fly as opposed to being stored. Next, we imagine we have $N$ particles distributed over the $M$ basis states and interpret $A_{ij}$ as the weight of particles arriving in state $|i\rangle $ on iteration $n+1$ per unit weight of a particle in state $|j\rangle $ on iteration $n$ and will regard the action of $A$ on a $\psi$ as causing a particle to jump from some $|j\rangle$ to some $|i\rangle$, carrying its current weight $\omega_j$, modified by $A_{ij}$, to state $|i\rangle$. To do this, we define the total weight leaving state $|j\rangle$ as
\begin{equation}
W_j=\sum_i A_{ij}
 \label{eq:weight_multiplier}
\end{equation}
and the transition probability from $|j\rangle$ to $|i\rangle$ as
\begin{equation}
T_{ij}=A_{ij}/W_j
\label{eq:transition_probability}
\end{equation}
Instead of always (i.e., with probability 1) moving weight $A_{ij}$ from state $|j\rangle$ to state $|i\rangle$, we will instead sample a $|i\rangle$ from $T_{ij}$ and multiply the transferred weight by $W_j$

For many Monte Carlo simulations, as is the case for the transfer matrix of the Ising model, the particle weights defining the eigenvector associated with the largest eigenvalue can be made all positive. The second eigenfunction however must be represented by some particles of negative weight and some particles of positive weight. For some jumps these negative and positive weights must at least partially cancel to maintain a correct estimation of the second eigenfunction. When $N \ll M$, as is typical, this cancellation does not occur often enough in a Monte Carlo simulation without proper design.
 
There are several ways to design the cancellation \cite{booth1}. For our Ising simulations, we promoted this cancellation by sorting the particles into state order (a state is represented by the bits of an integer) at the end of each iteration. Particles 1 and 2 are then sampled together according to the Arnow {\it et al.\/} scheme \cite{arnow}, then particles 3 and 4, and so forth. An ordered list means there are (typically) many nearby states $|i\rangle$ accessible from both particles $\ell$ and $\ell+1$ with nontrivial transition probabilities.

As the iteration progresses, the absolute value of the weights of some particles becomes very large, and those of some others, very small. As standard for Monte Carlo methods with weighted particles, we stochastically eliminated particles with weights of small magnitude and stochastically split those with large magnitudes. To do this we used a procedure called the comb \cite{comb1}.

The steps of the algorithm are: First, we initialize the weights of two vectors. For the Ising simulation, we selected $\omega_i'$ uniformly and randomly over the interval (0,1) and the $\omega_i''$ uniformly and randomly over the interval (-1,1). Then, for a fixed number of times we iterate. For each iteration we execute the jump procedure for each particle, place the particle list in state order,  effect cancellations, estimate the eigenvalues from (\ref{eq:eigenvalues}), update $\psi'$ and $\psi''$, and then comb. 

If $M$ is  sufficiently small so we can store all components of our vectors, sampling from the cumulative probability $C_i=\sum_{k=0}^i T_{kj}$ works well. If the number of states gets too large (e.g., $m>12$), then $C_i$ cannot be sampled directly because it cannot fit in the computer's memory.  In this case, we could just randomly pick from any state $|j\rangle$ any state $|i\rangle$ with probability $1/M$ instead of always picking a state $|i\rangle$ (i.e., with probability 1). The problem with this approach is that the $A_{ij}$ can have immense variation so that this simple sampling scheme is unlikely to work well as a Monte Carlo method. This situation is especially true for the Ising problem. A large part of such variations however can be removed by sampling the new state in stages and then sewing the stages together. 

To explain our sewing procedure, we will first assume that we can write any state $|i\rangle$ in our basis as a direct product of the states in a smaller basis, $
\left| i \right\rangle  = \left| {i_2 } \right\rangle \left| {i_1 } \right\rangle $. Instead of transferring weight
$W_j$ from state $|j\rangle$ to state $|i\rangle$ with probability $T_{ij}$, we will use the $a_{ij}$ that would apply to the smaller set of states and then make an appropriate weight correction. For the Ising model, $a_{ij}$ is the transfer matrix of a smaller lattice size.

For each smaller set of states, we rewrite the analogous transition probability from state $|j\rangle$ to state $|i\rangle$ as
\begin{equation}
t_{ij}=a_{ij}/w_j,
\label{eq:trans_mtrx}
\end{equation}
and the analogous weight multiplier as
\begin{equation}
w_j=\sum_k a_{kj}
\label{eq:wgt_mult}
\end{equation}
We thus will sample $|i_1\rangle$ and $|i_2\rangle$ from the probability function
\begin{equation}
  t(i_2,j_2)   t(i_1,j_1)
\label{eq:trans_prob}
\end{equation}
The weight correction $C_{ij}$, necessary to preserve the expected weight transfer from state $|j\rangle$ to $|i\rangle$, satisfies
\begin{equation}
A_{ij}=  C_{ij}  t(i_2,j_2)   t(i_1,j_1)
\label{eq:transfer_mtrx}
\end{equation}
Thus
\begin{equation}
C_{ij} = w_{j_1} w_{j_2} \frac{A_{ij}} {a(i_1,j_1) a(i_2,j_2) }
\label{eq:weight_correction}
\end{equation}

\begin{table*}[bht]
\caption{Our Monte Carlo power method estimates of $\lambda_1$ and $\lambda_2$ and their statistical errors for variously sized square Ising models. Each estimate was based on 20 independent simulations. Also given are the values computed via Onsager's exact results \cite{onsager}. \label{table:2}}
\begin{ruledtabular}
\begin{tabular}{lllll}
  Matrix Size & $\lambda_1$ (Onsager)& $\lambda_1$ & $\lambda_2$ (Onsager)& $\lambda_2$\\
 \hline
 $2^{16}\times 2^{16}$  & $2.93297\times 10^6$       & $2.93307 \pm 0.00008\times 10^6$
                                      & $2.79225\times 10^6$       & $2.79482 \pm 0.00010\times 10^6$\\
  $2^{32}\times 2^{32}$  & $8.39316\times 10^{12}$   & $8.39311\pm  0.00049\times 10^{12}$
                                      & $8.18959\times 10^{13}$   & $8.18807\pm  0.00061\times  10^{12}$\\
 $2^{48}\times 2^{48}$  & $2.41504\times 10^{19}$   & $2.41522 \pm 0.00019\times 10^{19}$
                                      & $2.37584\times 10^{19}$   & $2.37481 \pm 0.00054\times 10^{19}$ \\
\end{tabular}
\end{ruledtabular}
\end{table*}

The sewing method generalizes easily. For $k$ sets of states, (\ref{eq:transfer_mtrx})  and (\ref{eq:weight_correction})  become
\begin{equation}
A_{ij}=  C_{ij}  \prod_{n=1}^k t(i_n,j_n)
\end{equation}
with
\begin{equation}
C_{ij}=A_{ij} \prod_{n=1}^k\frac{w_{j_n}}{a(i_n,j_n)}
\end{equation}
For the Ising problem, we took for state $|i_1\rangle$ first $m/k$ bits of the integer representing $|i\rangle$; for $|i_2\rangle$, the second set; etc. The weight correction becomes
\begin{equation}
C_{ij} =\exp(\nu D_i)\prod_{n=1}^k w_{j_n}
\end{equation}
where apart form a factor of $\nu$, $D_i$ is the energy difference between calculating with the bits together and the bits separately. It is straightforward to calcualte.

Using this sewing algorithm for the sampling of states, we computed the first and second eigenvalues for $m=16$ to  $m=48$ Ising models by sewing 6 sets of 8 bits. We note that $2^{48}\approx 2.8\times 10^{14}$. The results are shown in Table~\ref{table:2}. To compute averages and standard errors for each size, we executed twenty independent Monte Carlo runs, each with 1 million particles per iteration (5 million for $m=48$), 500 iterations per run and used only the second half of the iterations in each run for the estimation process. $\nu=0.4406867935097715$ \cite{thompson,onsager}, the value at the critical temperature. Presented are 3 systems sizes. For each we give the values of $\lambda_1$ and $\lambda_2$ predicted from Onsager's expression and from our enhanced power method. We see that our Monte Carlo produced eigenvalues agree very well with Onsager's predictions. For our eigenvalue estimates, $R_1$ consisted of the states for which more than half of its $m$ bits were $0$'s and $R_2$ consisted of the states for which more than half of its $m$ bits were $1$'s.  More results and algorithmic detals will be given elsewhere \cite{booth3}.

We anticipate the sewing algorithm being applicable to other many-body problems defined on a lattice. Likely, these applications will require more sophisticated programming than for the Ising model. The ``sewing" method for the Ising model worked well as high as $m=60$, sewing together 6 sets of 10 bits. For $m>60$, our computer codes would need significant modification to implement a more flexible scalable procedure for representing a state configuration requiring more than a single computer word. We do not know how large an $m$ can be accommodated with a better computer program.
  
The transfer matrix of the Ising model is real, positive, asymmetric, and dense. How is our algorithm changed if a matrix lacks one or more of these properties? For simple test cases, we have successfully constructed deterministic procedures for matrices whose elements are complex valued. Also, in this context, we have had success for real asymmetric matrices whose eigenvalues are complex valued. Devising Monte Carlo algorithms for real, symmetric, sparse matrices has however received more of our attention \cite{gubernatis2}.  To find the eigenvalue of smallest size, if it not the one with the largest absolute value, one simply uses a shifted matrix, $A\rightarrow A-\sigma I$.

In closing, we believe that our new algorithm is accurate, easy to implement, and applicable to many other problems. Wider use of the algorithm will define more crisply its strengths and limitations than is possible by just the present application. The intent of the present application was benchmarking and not studying the scaling of the eigenvalues of the transfer matrix of the two-dimensional Ising model. Both deterministic and Monte Carlo power methods have been used for such studies. Deterministic methods \cite{richards93} have computed $\lambda_1$ and $\lambda_2$, while Monte Carlo methods \cite{examples}, just $\lambda_1$. The system sizes were considerably  smaller ($m\le 25$ deterministically and $m\le 21$ stochastically) than the largest size ($m=48$) presented here. This size should not be the largest accessible by our methods. All our calculations  were done on a single processor.

\begin{acknowledgments}
We thank M. E. Fisher for a helpful conversation. We gratefully acknowledge support of the U. S. Department of Energy through the LANL/LDRD program.
\end{acknowledgments}
 

\end{document}